%% file: skeleton.tex
\documentclass[a4paper,11pt]{article}
\usepackage{pos}

\usepackage{natbib}
\title{The Data Processor of the SPB2 Fluorescence Telescope: in flight performance}
 \ShortTitle{The DP of the SPB2 FT: in flight performance}

\author[a,b,1]{Valentina Scotti}
\author[b]{Antonio Anastasio} 
\author[b]{Alfonso Boiano} 
\author[c]{Francesco Cafagna}
\author[d,e]{Claudio Fornaro}
\author[b]{Vincenzo Masone} 
\author*[a,b]{Marco Mese} 
\author[b]{Giuseppe Osteria} 
\author[a,b]{Francesco Perfetto}
\author[b]{Gennaro Tortone} 
\author[b]{Antonio Vanzanella} 

\affiliation[a]{Dipartimento di Fisica "E. Pancini", "Università degli Studi di Napoli Federico II",\\
  Complesso Universitario di Monte SantAngelo, Napoli, Italy}

\affiliation[b]{Istituto Nazionale di Fisica Nucleare, Sezione di Napoli,\\
  Complesso Universitario di Monte SantAngelo, Napoli, Italy}
  
\affiliation[c]{Istituto Nazionale di Fisica Nucleare, Sezione di Bari,\\
  Via E. Orabona 4, Bari, Italy}

  \affiliation[d]{Università Telematica Internazionale Uninettuno,\\
  Corso Vittorio Emanuele II 39, Roma, Italy}

  \affiliation[c]{Istituto Nazionale di Fisica Nucleare, Sezione di Roma Tor Vergata,\\
  Via della Ricerca Scientifica 1, Roma, Italy}

  \note{Corresponding author}
  \renewcommand{\printHeadAuthors}{V. Scotti et al.}
\onbehalf{for the JEM-EUSO Collaboration\\[-1mm]{\normalsize \normalfont (a complete list of authors can be found at the end of the proceedings)}}

\emailAdd{scottiv@na.infn.it}

\abstract{EUSO-SPB2 (Extreme Universe Space Observatory on a Super Pressure Balloon II) is a precursor mission for a future space observatory for multi-messenger astrophysics, planned to be launched in Spring 2023 with a flight duration target of 100 days. The Fluorescence Telescope (FT) hosted on board is designed to detect Ultra High Energy Cosmic Rays via the UV fluorescence emission of the Extensive Air Showers in the atmosphere.
The Data Processor (DP) of the FT is the component of the electronics system that performs data management and instrument control for the telescope. The DP controls front-end electronics, tags events with arrival time and payload position through a GPS system, provides signals for time synchronization of the event and measures the live and dead time of the telescope. Furthermore, it manages mass memory for data storage, performs housekeeping monitoring, and controls the power-on and power-off sequences. Finally, the data processor combines the data from the PDMs and onboard differential GPS and prioritizes data for download.
The long duration of the flight poses strict requirements on electronics and data handling. The operations at high altitude in an unpressurized environment represent a technological challenge for heat dissipation.
This contribution will provide an overview of the innovative elements developed and the results of the integration and field test campaigns. We will also present some preliminary analysis of the performance during the flight.
}

\ConferenceLogo{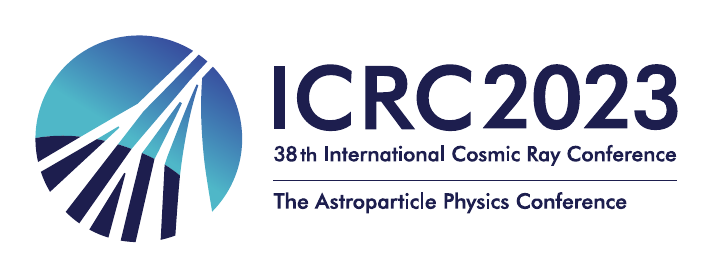}

\FullConference{%
38th International Cosmic Ray Conference (ICRC2023)\\
  26 July - 3 August, 2023\\
  Nagoya, Japan}

\begin{document}
\maketitle

\section{Introduction}
\label{sec:introduction}

EUSO-SPB2 is a precursor mission for forthcoming space observatories for multi-messenger astrophysics. The innovative approach of EUSO-SPB2 is to use the Earth as a giant particle detector, studying fluorescence and Cherenkov light from the edge of space. 

The primary objective of this mission is to detect fluorescence light generated in the Earth's atmosphere by Ultra High Energy Cosmic Rays (UHECR) from above, to confirm the expectations from ground observations \cite{auger,ta}.  In addition, this mission explores new areas such as detecting Cherenkov light emission from air showers and measuring the background of up-going tau decays from cosmogenic neutrinos \cite{poemma}.

Drawing upon the rich experience gleaned from its predecessor, EUSO-SPB1 \cite{euso-spb1}, EUSO-SPB2, paves the path towards the prestigious Probe of Extreme Energy Multi-Messenger Astrophysics (POEMMA) space mission \cite{poemma}. POEMMA is designed to unravel the very essence and origins of Extreme Energy Cosmic Rays (EECR) while uncovering the elusive emissions of astrophysical neutrinos. 

EUSO-SPB2 has been launched in May 2023 from Wanaka in New Zealand. A description of the mission is given in \cite{euso-spb2}. In this paper, we present the concept of the Data Processor (DP) system designed to manage the Fluorescence Telescope of EUSO-SPB2 while respecting the stringent constraints posed by a balloon-borne experiment.

\section{The EUSO-SPB2 fluorescence telescope}

EUSO-SPB2 hosted onboard two independent telescopes, a fluorescence detector (FT) and a Cherenkov detector (CT), in order to be a scientific and technical sub-orbital altitude precursor for the POEMMA. A complete description of the two instruments can be found respectively in \cite{fluorescence} and \cite{cherenkov}, here we will briefly report only on the details relevant to the functioning of the Data Processing system. 

The fluorescence detector has been built on the experience of EUSO-SPB1 to detect Ultra-High Energy Cosmic Rays via the UV fluorescence emission of the extensive air showers generated by UHECR in the atmosphere{UV}. The main difference with respect to EUSO-SPB1 is that in EUSO-SPB2 the light is focused on the camera through a Schmidt optical system, instead of Fresnel lenses. In addition, the focal surface of the FT consists of 3 Photo Detector Modules (PDM), each composed of 36 MAPMT Hamamatsu R11265, for a total of 6912 pixels, compared to 2304 of EUSO-SPB1. The Field of View of each PDM is $12^{\circ}\times12^{\circ}$. The signals from MaPMT are readout by a specific ASIC followed by multiple trigger levels to filter out the noise and identify events of interest. This system has single photoelectron counting capability and a time unit of $1.05 \mu s$. The FT points to nadir and the expected trigger rate from EASs according to simulations is around 0.2/hour for a trigger threshold of $10^{18}$eV. 

The telescope is completed by several ancillary devices:
\begin{itemize}
    \item Light sensor, called EMON, located inside the telescope, to verify darkness conditions during operation and testing.
    \item Health LED Systems to check camera health and response.
    \item A dual-band InfraRed camera to monitor the presence of clouds and to measure cloud-top temperatures only for the fluorescence telescope.
\end{itemize}

\section{The Data Processing system}

The architecture of the Data Processing system designed for the EUSO-SPB2 FT is an evolution of the one developed for the fluorescence telescope of the EUSO-SPB1 \cite{dp}.

\begin{figure*}[bt]
\centerline{\includegraphics[width=\textwidth]{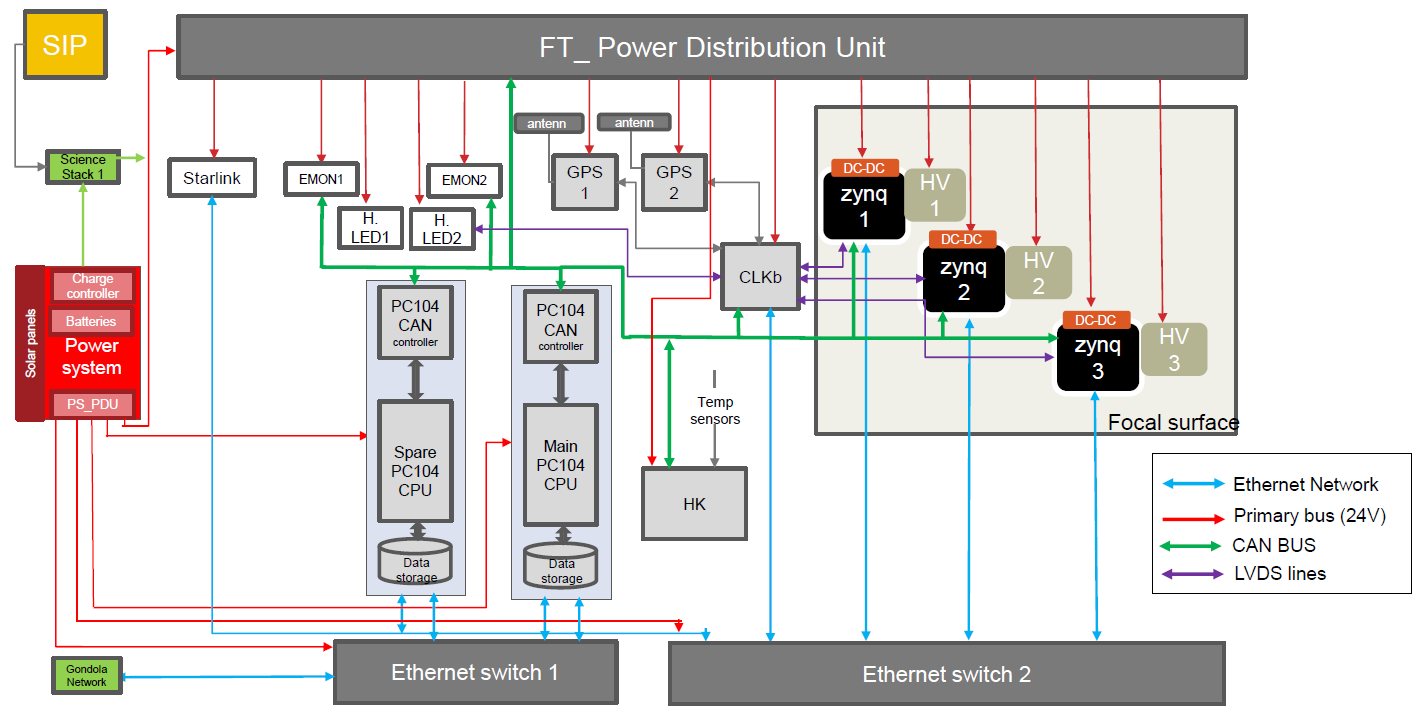}}
\caption{Block diagram of the Data Processing system and its connections with the rest of the instrument for the fluorescence telescope of the EUSO-SPB2.}
\label{fig:blockdiag}
\end{figure*}

The block diagram of the DP of the fluorescence telescope and its connections with the rest of the instrument is shown in Fig.~\ref{fig:blockdiag}. 
In the following paragraphs, we will describe the Data Processing system architecture and its performance during the flight.

\subsection{Requirements}

The EUSO-SPB2 FT was designed to collect data during moonless nights to obtain the best contrast for the signatures of scientific interest. The DP links the telescope with the Gondola system. The DP is a complex system that includes most of the digital electronics of the instrument which allows a comprehensive control, configuration, monitoring, and operation of the telescope throughout the commissioning phase, test campaigns, and the entirety of the flight mission.

Besides being the interface with the flight computer, the data processing system was designed to perform the following tasks:
\begin{itemize}
        \item Main interface with Flight Computer (GCC) telemetry system
        \item Data selection/compression and transmission to Flight Computer (SIP)
        \item Power ON/OFF the whole instrument
        \item Define Telescope operation mode (Day, Night, D-N-D transitions)
        \item Configure the Front End electronics
        \item Start/Stop of the data acquisition and calibration procedures
        \item Tag events with GPS time and GPS position
        \item Synchronization of the data acquisition
        \item Manage trigger signals (L1 by the 3 PDM Boards, external, GPS, by CPU command, etc) 
        \item Manages mass memory for data storage
        \item Monitor/Control/DAQ of some Ancillary Devices
        \item Monitor voltages, current, and temperatures (LVPSs, boards, FPGAs)
\end{itemize}

The main challenges in designing and building the system are due to the physical characteristics of the signal and to the harsh environment in which the DP has to operate.

At the Peak energy sensitivity for the FT (around 6 EeV) the expected event rate is around 1 event every 13 hours. For an assumed 80-day flight with a 15\% duty cycle, the total number of events is around 22 events, but the trigger rate of the FT was estimated, and it was confirmed during flight operation, to be between 5 and 10 Hz, so it’s necessary to distinguish a few cosmic rays events among the background. 

Furthermore, since the target flight duration for the NASA super pressure program is 100 days, the requirements on electronics and data handling are quite severe. Finally, the system operated at high altitude in an unpressurized environment, which introduces a technological challenge for heat dissipation.

In addition to power and mass budget restrictions on the balloon payload, there is also a limited telemetry budget of 1GB per day. For this reason, there is the need to prioritize data for downloading on board, besides the necessity to only record and transfer high-quality events. Everything must be done on board with minimal intervention from the ground.

\section{The subsystems}

The block diagram (Fig. \ref{fig:blockdiag}) shows that the data processing system is composed of several subsystems. The whole system is capable to acquire 7000 channels without exceeding the mass and power budget. 

Different tasks are performed by different subsystems. All the subsystems, together with the low-voltage power supply modules, are hosted in a DP box (customized Eurocard chassis) equipped with a cooling plate to dissipate heat (Fig. \ref{fig:structure}):
\begin{figure}[bt]
\centerline{\includegraphics[width=.85\linewidth]{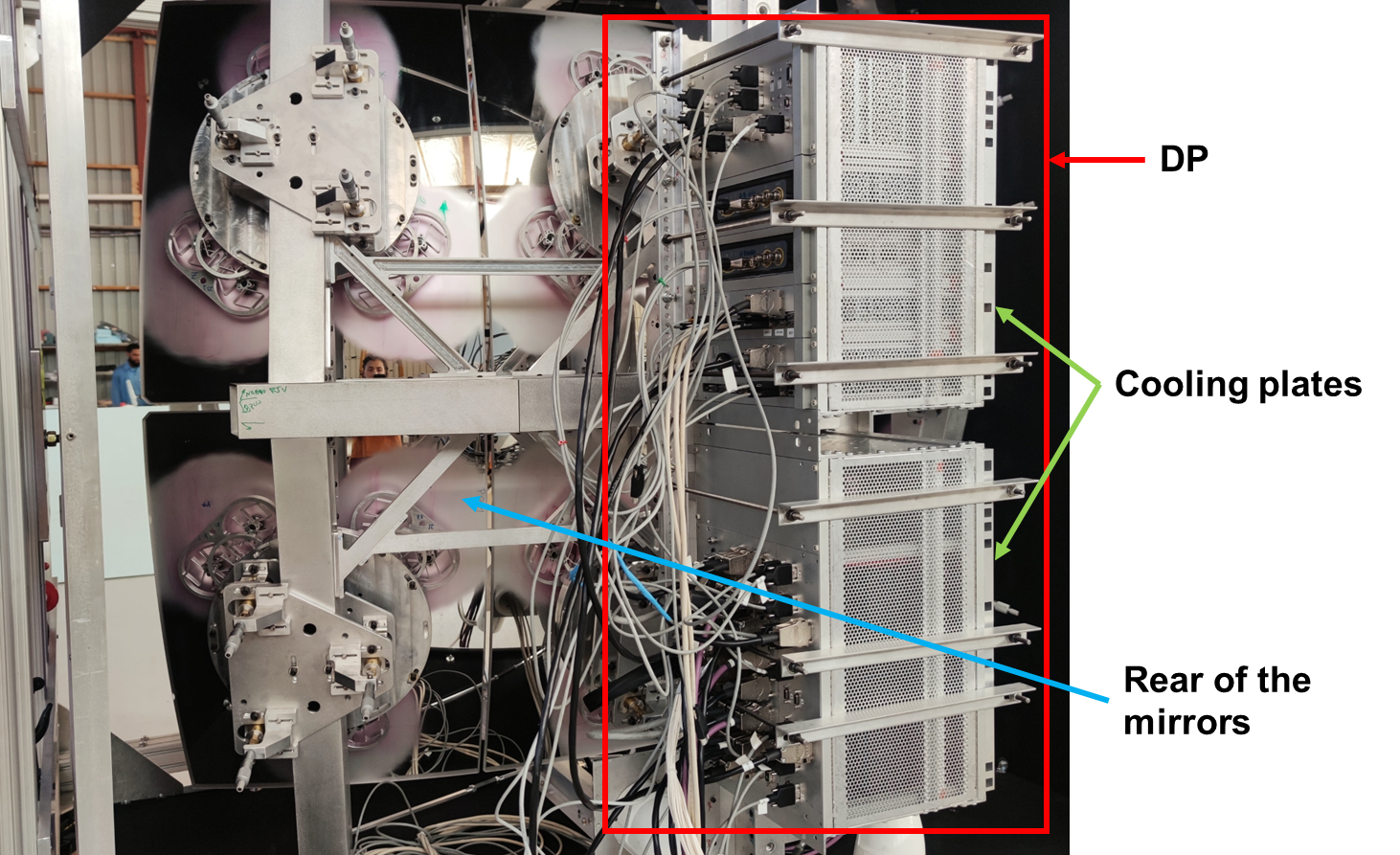}}
\caption{The DP box, hosting all the Data Processing subsystems, mounted on the back of the telescope structure. The cooling plate is on the back of the box.}
\label{fig:structure}
\end{figure}
\begin{itemize}
    \item CPU with Hot and Cold redundancy PCI/104 single board computer Core i7 3517UE + 5 SATA disks for data storage;
    \item HK board based on STM32 microcontroller and TIBBO EM2000;
    \item Clock board \cite{clock} based on Xilinx XC7Z SoC:  manages trigger and data synchronization and also acts as an interface with the GPS receivers;
    \item two GPS receivers Trimble BX992;
    \item Ethernet switches;
    \item Solid State Power Controller.
\end{itemize}

We choose to use mostly COTS devices because reliability in a hostile environment is one of the requirements. The electronic components are all selected to operate in an extended temperature range.

For the most fundamental component, the CPU, we have hot and cold redundancy, hence the choice to use CAN and Ethernet protocols which guarantee us in case one CPU fails.

Thanks to the switch redundancy in the FT DP, a StarLink Module could be installed in Wanaka. The DP guaranteed the network separation requested by CSBF and had spare PDU (Power Distribution Unit) channels available. Starlink is a low Earth orbit satellite constellation that delivers high-speed, low-latency internet, the one chosen by CSBF featured up to 220 Mbps in download. 

\section{The software}

The DP functions as an interface, forging a critical link between the detector and the telemetry blocks, while also facilitating seamless interaction with the end users. Instrument operations and monitoring were shared between collaborators in the US, Japan, and Europe to provide, during local daytime, control, and monitoring of the telescopes 24/7, similarly to the on-ground control that was done for EUSO-SPB1 \cite{software}.  

The CPU software can be divided into two categories:
\begin{itemize}
    \item the control software 
    \item the data handling software
\end{itemize}

The control software exhibits a high level of flexibility, crafted to adapt to various phases encompassing both commissioning and the flight mission. Notably, this software allowed to control and monitor the detector, while also ensuring comprehensive data storage, encompassing scientific data, housekeeping records, and log information. The design enables full user control via telemetry, enabling real-time command and supervision. Significantly, during the flight mission, it is imperative to maintain continuous surveillance of the instrument's status. Such vigilant monitoring is essential to discern the opportune moments for initiating or halting measurements. The software's adeptness in facilitating constant status checks proved paramount in making decisive determinations, enhancing the precision and efficiency of the overall mission operations.

The data handling software manages science data acquisition from the CLK and Zynq boards. The acquisition software initializes and configures the subsystems, monitors the connections with the CPU, and verifies the behavior of the subsystems. It can perform several types of acquisitions, such as recording externally triggered events or internally triggered events. Finally, it manages the prioritization folder structure for down-link to cope with the maximum amount of telemetry available each day (1 GB), evaluated when the use of Starlink was not an option.

Science Stack data were available in real time and were used to monitor and control the status of the telescope. In particular, before starting the power-on or power-off sequence, it was necessary to monitor:
\begin{itemize}
    \item Status of the power system;
    \item Temperatures of critical points of the telescope (Batteries, mirror, focal surface, electronics, etc.);
    \item Light level of the EMON (Day/Night transition, Night/Day transition);
    \item Status of the aperture shutter (Open/Close).
\end{itemize}

\section{In-flight performance}

The instrument was launched with all but HV turned on, the communication of the DP with the Zynq Boards was verified throughout the ascent. The Housekeeping data was transmitted immediately via Starlink and quickly viewed in Grafana. DP system was stable and continuously acquired data during the dark periods and transmitted data to the ground (after compression) 24 hours a day during the flight until the payload hit the ocean.

The CPU worked at all times without any flaws and never needed a reboot. The system acquired data for a total of about 3 hours, and 56 GB of data was recorded on the disk. Thanks to Starlink's high throughput 39 GB of the acquired data were transmitted to the ground together with telemetry data (voltages, currents, temperatures, ...) as well as data from UCIRC and Cherenkov Telescope before the termination of the flight.

To evaluate the performance of the DP system, we studied the dead time and live time measured by the CLK-Board. The mean dead time was about 1.1 ms. The live time changes according to the trigger algorithm used, because lower trigger rates cause an increase in the live time, the mean acquisition rate was 7 Hz.

\begin{figure}[bt]
\centerline{\includegraphics[width=.8\linewidth]{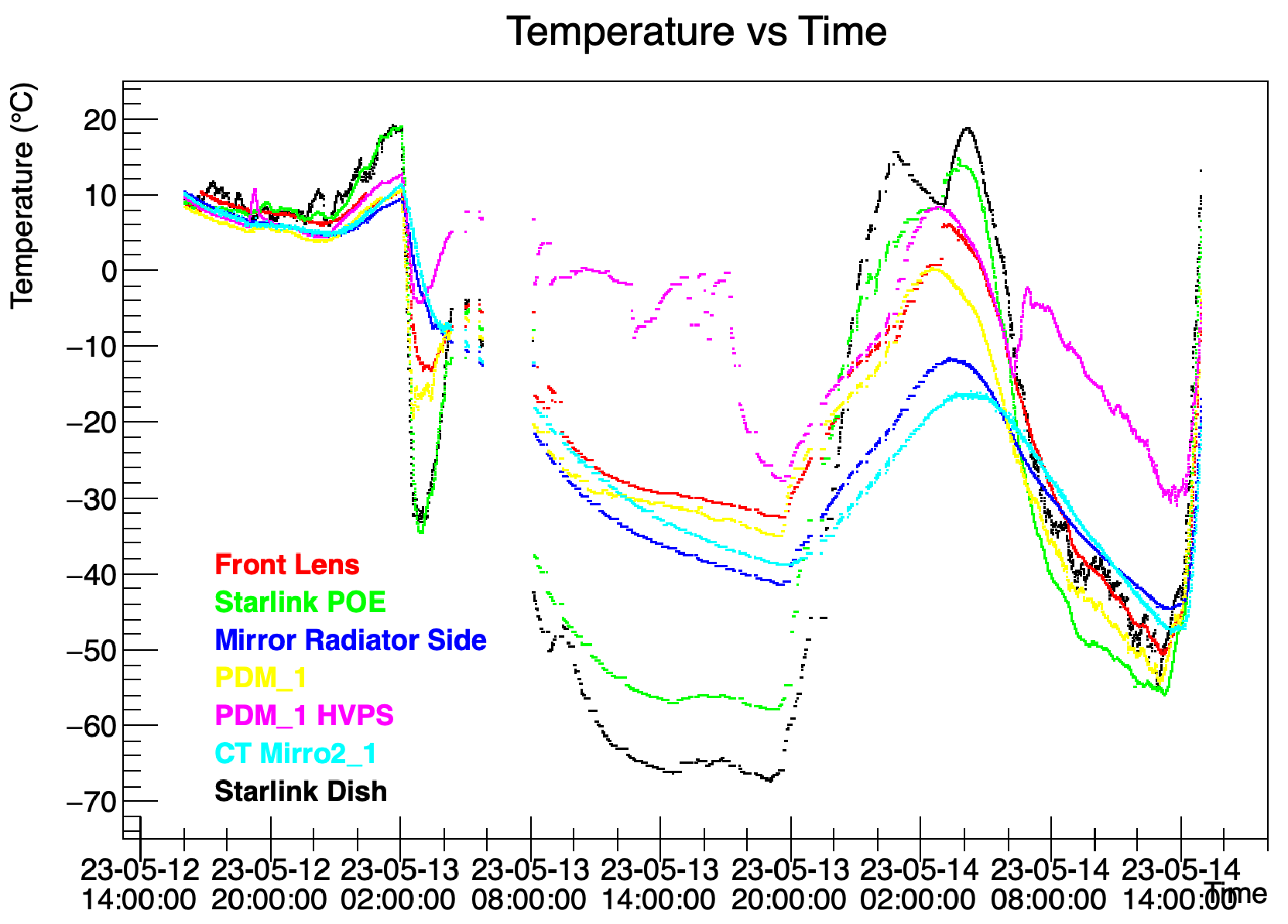}}
\caption{The temperature in several points of the instruments monitored by the DP.}
\label{fig:temp}
\end{figure}

The Data Processor system demonstrated its capability by providing complete control of the instrument throughout the entire flight mission, enabling seamless data-taking operations. All constituent subsystems of the DP operated as per their intended specifications. According to telemetry data analysis, the DP exhibited exceptional performance, continuously transmitting compressed data to the ground segment, up until the point of reaching the ocean surface. A plot of the Housekeeping data relative to the temperatures in several points of the instrument is shown in Figure \ref{fig:temp}. 
This uninterrupted data transmission exemplified the system's robustness in efficiently collecting, selecting, and transmitting data, thus ensuring a successful and productive mission.
			
The system's reliability was fully demonstrated through flawless data acquisition processes. The procedures devised for data acquisition and storage were so meticulously optimized that the acquisition speed was primarily constrained by the total data transfer capacity to the ground. Such performance underscored the system's efficacy and its capability to handle the demanding data requirements with efficiency.

\section{Conclusions}
EUSO-SPB2 is a 2-telescope instrument to study UHECR, Cherenkov, and tau neutrino backgrounds which has been launched in May 2023. During the flight, the Data Processing system of the Fluorescence Telescope worked flawlessly and allowed us to manage the instrument and download a good part of the total data acquired. 

\section*{Acknowledgment}

The authors acknowledge the support by NASA awards 11-APRA-0058, 16-APROBES16-0023, 17-APRA17-0066, NNX17AJ82G, NNX13AH54G, 80NSSC18K0246, 80NSSC18K0473, 80NSSC19K0626, 80NSSC18K0464, 80NSSC22K1488, 80NSSC19K0627 and 80NSSC22K0426, the French space agency CNES, National Science Centre in Poland grant n. 2017/27/B/ST9/02162, and by ASI-INFN agreement n. 2021-8-HH.0 and its amendments. This research used resources of the US National Energy Research Scientific Computing Center (NERSC), the DOE Science User Facility operated under Contract No. DE-AC02-05CH11231. We acknowledge the NASA BPO and CSBF staffs for their extensive support. We also acknowledge the invaluable contributions of the administrative and technical staffs at our home institutions.

\input{JEM-EUSO_Authors_July2023.tex}

%
%
%

\end{document}

%% file: JEM-EUSO_Authors_July2023.tex
\newpage
{\Large\bf Full Authors list: The JEM-EUSO Collaboration\\}

\begin{sloppypar}
{\small \noindent
S.~Abe$^{ff}$, 
J.H.~Adams Jr.$^{ld}$, 
D.~Allard$^{cb}$,
P.~Alldredge$^{ld}$,
R.~Aloisio$^{ep}$,
L.~Anchordoqui$^{le}$,
A.~Anzalone$^{ed,eh}$, 
E.~Arnone$^{ek,el}$,
M.~Bagheri$^{lh}$,
B.~Baret$^{cb}$
% ...
}
\end{sloppypar}
\vspace*{.3cm}

{ \footnotesize
\noindent
% ...
$^{cb}$ Universit\'e de Paris, CNRS, AstroParticule et Cosmologie, F-75013 Paris, France\\
% ...
$^{ed}$ Istituto Nazionale di Fisica Nucleare - Sezione di Catania, Italy\\
% ...
$^{eh}$ INAF - Istituto di Astrofisica Spaziale e Fisica Cosmica di Palermo, Italy\\
% ...
$^{ek}$ Istituto Nazionale di Fisica Nucleare - Sezione di Torino, Italy\\
$^{el}$ Dipartimento di Fisica, Universit\`a di Torino, Italy\\
% ...
$^{ep}$ Gran Sasso Science Institute, L'Aquila, Italy\\
% ...
$^{ff}$ Nihon University Chiyoda, Tokyo, Japan\\ 
% ...
$^{ld}$ University of Alabama in Huntsville, Huntsville, AL, USA\\
$^{le}$ Lehman College, City University of New York (CUNY), NY, USA\\
% ...
$^{lh}$ Georgia Institute of Technology, USA\\
% ...
}